\documentclass[12pt,a4paper]{article}
\usepackage[utf8]{inputenc}
\usepackage{amsmath}
\usepackage{amssymb}
\usepackage{graphicx}
\usepackage{subfigure}
\usepackage[affil-it]{authblk}
\usepackage{etoolbox}
\usepackage{lmodern}

\makeatletter
\patchcmd{\@maketitle}{\LARGE \@title}{\fontsize{16}{19.2}\selectfont\@title}{}{}
\makeatother

\author[1,2]{Rafael A. Bizao}
\author[1,3]{Leonardo D. Machado}
\author[1,4]{Jose M. de Sousa}
\author[2,5,6,*]{Nicola M. Pugno}
\author[1,+]{Douglas S. Galvao}
\affil[1]{Instituto de F\'isica Gleb Wataghin, Universidade Estadual de Campinas, 13083-970, Campinas, SP, Brazil.}
\affil[2]{Department of Civil, Environmental and Mechanical Engineering, Laboratory of Bio-Inspired and Graphene Nanomechanics, University of Trento, via Mesiano, 77, 38123 Trento, Italy.}
\affil[3]{Departamento de F\'isica Te\'orica e Experimental, Universidade Federal do Rio Grande do Norte, Natal-RN 59072-970, Brazil.}
\affil[4]{Departamento de F\'isica, Universidade Federal do Piau\'i, Teresina, Piau\'i, 64049-550, Brazil.}
\affil[5]{Italian Space Agency, Via del Politecnico snc, 00133 Rome, Italy.}
\affil[6]{School of Engineering and Materials Science, Queen Mary University of London, Mile End Road, London E1 4NS, United Kingdom.}
\affil[*]{nicola.pugno@unitn.it}
\affil[+]{galvao@ifi.unicamp.br}
\date{}

\title{Scale Effects on the Ballistic Penetration of Graphene Sheets}

\begin{document}
\maketitle

\begin{abstract}
Carbon nanostructures are promising ballistic protection materials, due to their low density and excellent mechanical properties. Recent experimental and computational investigations on the behavior of graphene under impact conditions revealed exceptional energy absorption properties as well. However, the reported numerical and experimental values differ by an order of magnitude. In this work, we combined numerical and analytical modeling to address this issue. In the numerical part, we employed reactive molecular dynamics to carry out ballistic tests on single and double-layered graphene sheets. We used velocity values within the range tested in experiments. Our numerical and the experimental results were used to determine parameters for a scaling law, which is in good agreement with all experimental and simulation results. We find that the specific penetration energy decreases as the number of layers ($N$) increases, from $\sim25$ MJ/kg for $N=1$ to $\sim0.26$ MJ/kg as $N \to\infty$. These scale effects explain the apparent discrepancy between simulations and experiments.
\end{abstract}


\section*{Introduction}

The combination of very high Young's modulus ($1$ TPa), ultimate strength ($130$ GPa), and low density values ($\approx 2200$ kg.m$^{-3}$) makes graphene an ideal candidate material for ballistic protection applications \cite{pugno2007nanotube}. However, the rapid strain increase found in these applications can lead to unexpected behavior. For instance, experiments in this regime revealed unzipping of carbon nanotubes (CNTs) into nanoribbons  \cite{ozden2014unzipping}. While the high-strain-rate behavior of CNTs, either isolated \cite{mylvaganam2007ballistic,coluci2007mechanical} or in composites \cite{pandya2012ballistic, laurenzi2013experimental, obradovic2015dynamic,signetti2014evidence}, has been studied for years,  investigations on graphene mainly date from 2014  \cite{Lee2014,eller2015hypervelocity,zhang2015finite,wetzel2015theoretical,yoon2016atomistic, haque2016molecular,xia2016failure}. Of particular interest is the study by Lee \textit{et al.} \cite{Lee2014}, in which silica spheres were shot at multilayered graphene sheets. Exceptional energy absorption capabilities were found: the specific penetration energy of graphene was ten times greater than that of macroscopic steel. This was due in part to the impact energy being dissipated over an area much larger than that of the projectile cross-section. 

Follow-up molecular dynamics (MD) studies elucidated the atomistic structures formed during penetration of graphene monolayers and the role played by defects  \cite{yoon2016atomistic}, determined the propagation velocity of the impact-induced stress wave \cite{haque2016molecular}, and studied the failure mechanism of the graphene sheets \cite{xia2016failure}. These simulations also revealed extremely high specific energy penetration values, an order of magnitude greater than those measured in experiments. Up to now this large discrepancy between theory and experiment has remained unexplained.  In this work, we combined fully atomistic reactive MD simulations and analytical modeling to address this issue.

\section*{Results and Discussions}

\subsection*{Simulated ballistic tests}

In the MD part of our study, we shot metallic projectiles at single and bi-layer graphene sheets. We have considered different projectile velocities and impact angles, as well as sheets of different dimensions (up to 400,000 atoms). As further discussed below, we also obtained MD specific penetration energy values that are one order of magnitude larger than those from experiments, but the difference decreased for the bi-layer case. From these results, we were able to extract parameters to apply in a scaling law proposed by Pugno \cite{Pugno2007}. Our analytical model fits very well all existing results for graphene, and suggests an asymptotic value of $\sim 0.26$ MJ/kg for graphene in the macroscopic limit (when the number of layers (N) tends to infinity).

\begin{figure}
\centerline{\includegraphics[width=0.9\linewidth]{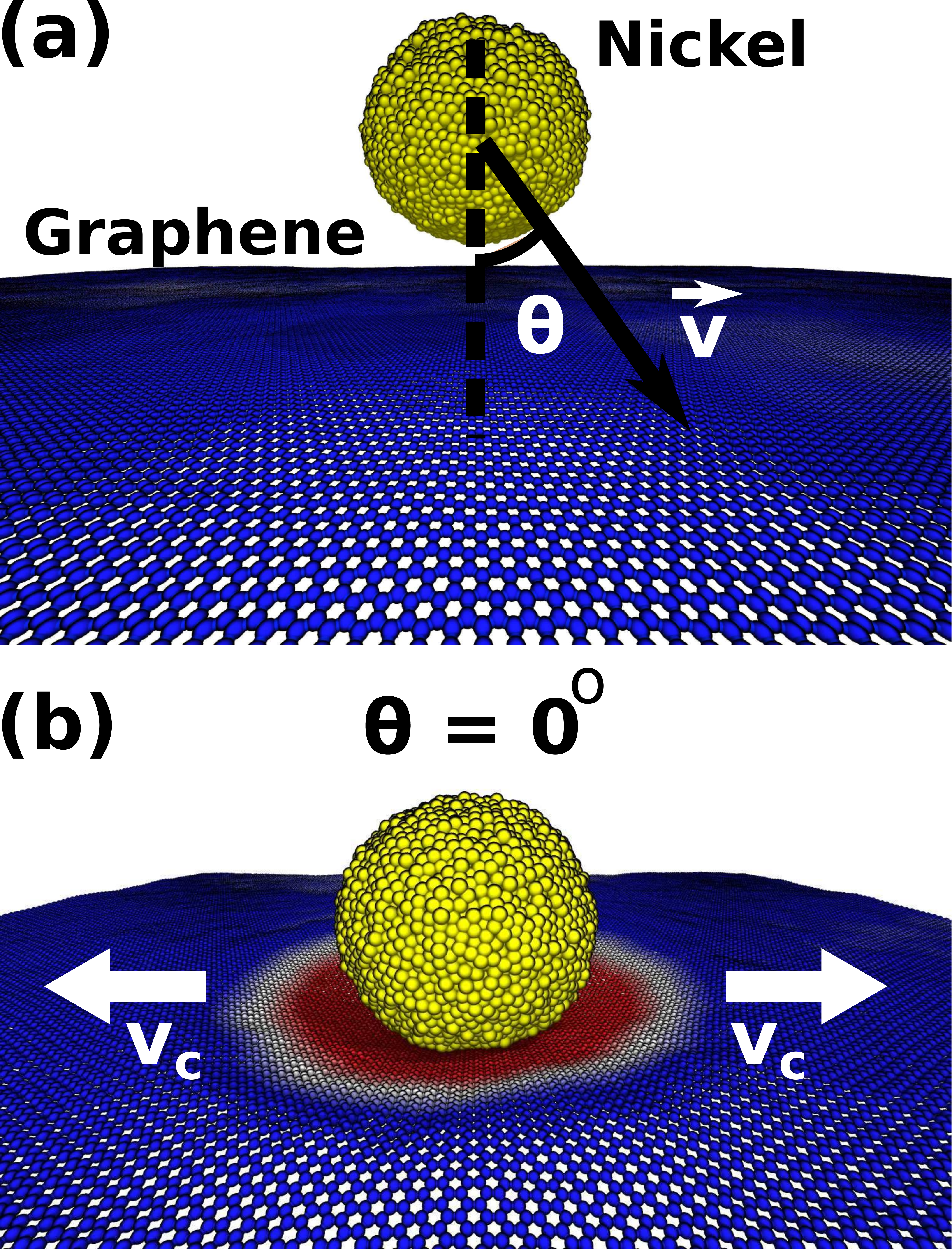}}
\caption{(a) Setup employed in the fully atomistic molecular dynamics (MD) simulations. We shot a nickel particle against graphene sheets, at different velocity \textbf{v} and angle $\theta$ values. (b) MD snapshot from a case with $\theta=0^{\circ}$ and $v=900$ m/s. The ballistic impact generates an elastic deformation wave that propagates with velocity $v_c$ over an area much larger than the particle dimensions.
\label{representacao}}
\end{figure}

A typical setup used in our ballistic tests is presented in figure \ref{representacao}a. The considered graphene targets are periodic along the planar directions, and ranged from $20$ nm $\times$ $20$ nm ($30,000$ atoms) up to $100$ nm $\times$ $100$ nm ($385,000$ atoms). For the smallest cases, we also considered bi-layer structures. For all simulations we used a spherical ($r \sim 3.5$ nm) nickel particle as projectile. Different \textbf{v} and $\theta$ values were considered (see Fig. \ref{representacao}a). Detailed information regarding the simulations can be found in the Methods section.

\begin{figure}
\centerline{\includegraphics[width=0.9\linewidth]{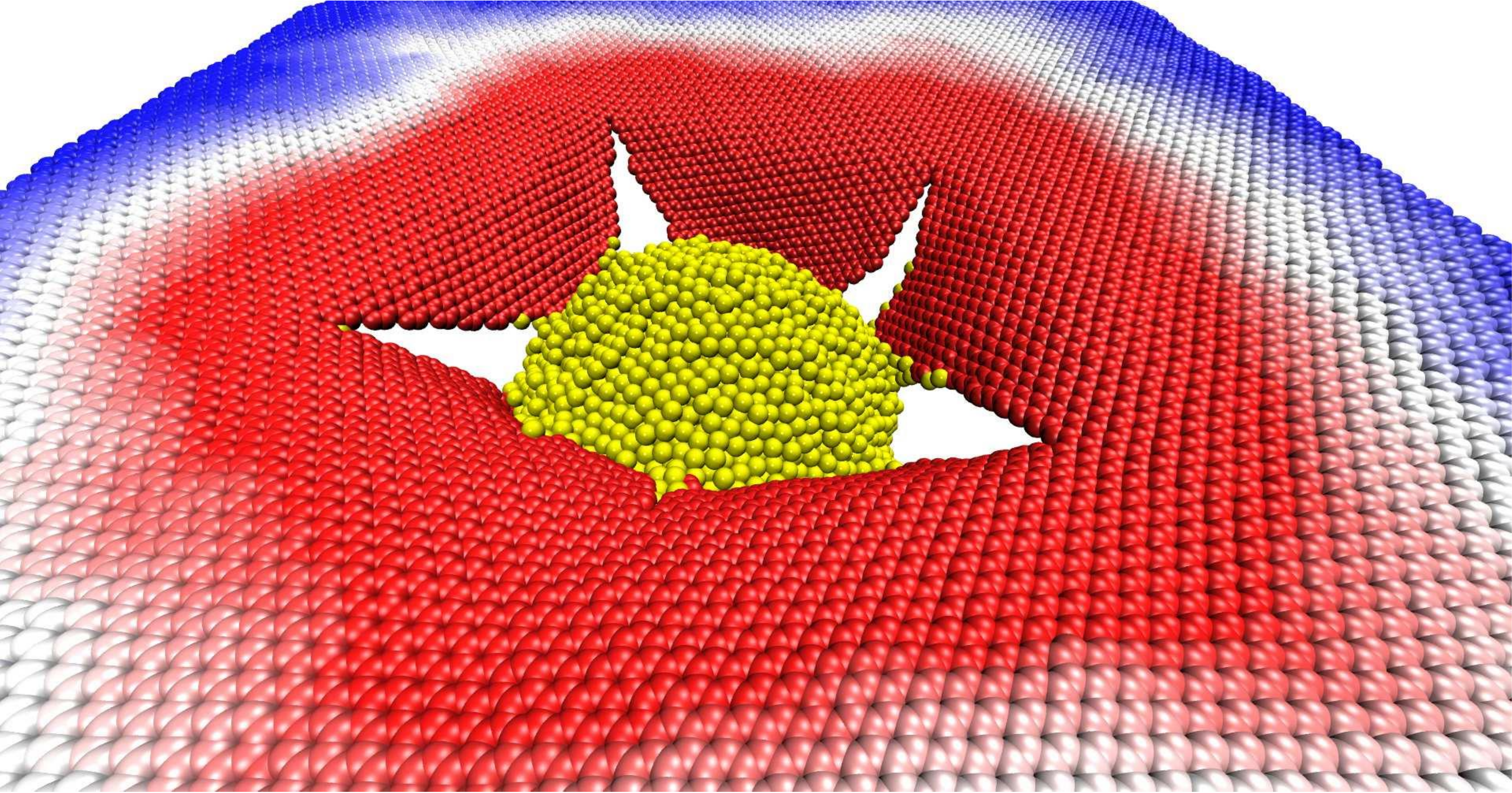}}
\caption{MD snapshot from a case of $\theta=0^{\circ}$ and $v=900$ m/s showing the fractured graphene sheet after the ballistic impact.
\label{representacao2}}
\end{figure}

\begin{figure}
\centerline{\includegraphics[width=0.6\linewidth]{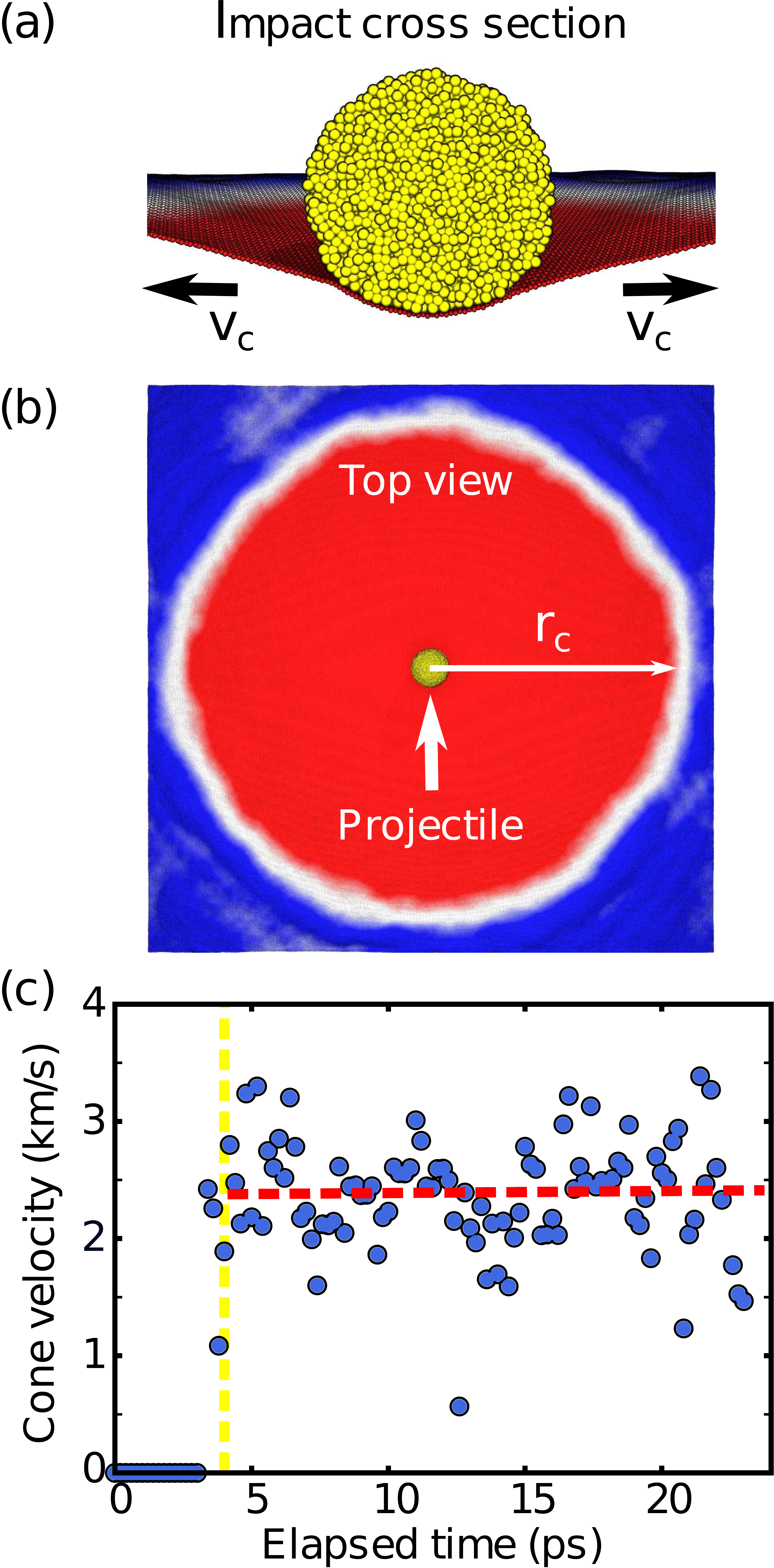}}
\caption{Results obtained for $\theta=0^{\circ}$ and $v=900$ m/s. (a) Impact cross sectional view, showing a graphene sheet deformed into a conical shape. (b) Top view of an impact. Observe that the deformation cone radius ($r_c$) is far larger than the projetile cross-section value. (c) Instantaneous cone velocity values. The linear fit (red line) suggests that, considering error bar fluctuations, the generated conical shape propagates at constant velocity. The points considered in the fit are to the right of the yellow line (impact time).}
\label{cone}
\end{figure}

In Figure \ref{representacao}b and Figure \ref{representacao2} we present MD snapshots for the case of $\theta=0^{\circ}$ and $v=900$ m/s. In Figures \ref{representacao}, \ref{representacao2} and \ref{cone}, graphene atoms are colored according to their z (height) coordinate values: positive values are in blue and negative ones in red. After impact, the generated elastic deformation wave propagates radially outwards with velocity $v_c$ - see figure \ref{representacao}b. In agreement with the report by Lee \textit{et al.} \cite{Lee2014}, we observed deformation areas far larger than the projectile cross-section (figure \ref{cone}b). Our typical fracture patterns are also consistent with experimental results \cite{Lee2014}. For a better visualization of the whole process see videos in the Supplementary Information.

From our MD trajectories we can analyze in detail the onset and propagation of the impact-generated elastic deformation wave. Inspection of the cross-sectional view of an impact event (Fig. \ref{cone}a) reveals that graphene stretches to accommodate the incoming projectile into a cone shape. Lee \textit{et al.} \cite{Lee2014} reached the same conclusion from their experiments and estimated, using the formula proposed by Phoenix and Porwal \cite{phoenix2003new}, a velocity of $v_c=2560$ m/s for an impact velocity of $900$ m/s. From our MD trajectories we can not only calculate average cone velocities, but also their time evolution. In our analysis, atoms that moved 12 \textup{\AA}  down from their initial position were assumed inside the cone. The first atoms to cross this threshold were considered at the impact center, and for every MD snapshot frame we calculated the distance from this center to the farthest atom in the cone, $r_c$ (see Fig. \ref{cone}b). If the time between adjacent frames is $\Delta t$ and the cone radius increased by $\Delta r_c$ in this interval, the instantaneous velocity can be calculated by using $v_c= \Delta r_c / \Delta t$.

Results of this analysis are presented in Figure \ref{cone}c, where the red dotted line is a linear fit of the data. More details are discussed in the Supplementary Information. For the case presented in Figure \ref{cone}c, we obtained a cone acceleration of $0.0017 \pm 0.0095$ km/s$^2$. Near zero acceleration values were also observed for other impact velocities, indicating that the cones propagated with constant velocity for all the analyzed events. For an impact at 900 m/s, we found $v_c = 2.37 \pm 0.14$ km/s, a value rather close to the estimation by Lee \textit{et al.} \cite{Lee2014}. Graphs for other impact velocities are also presented in the Supplementary Information. For an impact at 600 (1100) m/s we obtained average cone velocities of $1.99 \pm 0.15$ ($2.64 \pm 0.10$) km/s, values that are again close to those estimated by Lee \textit{et al.}, 1.95 (2.92) km/s \cite{Lee2014}. It should be remarked that Haque \textit{et al.} \cite{haque2016molecular} also found constant $v_c$ values in their MD simulations. However, under the higher velocity conditions they used, the obtained $v_c$ values were $\sim 35\%$ lower than those obtained by employing the formula by Phoenix and Porwal \cite{phoenix2003new}, suggesting a limit of validity for this expression.

In order to contrast our results against other theoretical \cite{yoon2016atomistic,haque2016molecular, xia2016failure} and experimental \cite{Lee2014} reports, we normalized the energy absorbed by the graphene mass within the projectile cross-sectional area, obtaining the specific penetration energy. This comparison is presented in table \ref{tab}. Note that the values attributed to Haque \textit{et al.} \cite{haque2016molecular} were calculated from data provided in their manuscript - see the Methods section for details. Currently reported numerical values are an order of magnitude larger than experimental ones, although the difference decreased for the considered bilayer systems. It is important to remark that direct comparison between numerical (up to now single and bilayer systems) and experimental (up to now from 30 up to 300 layers) results is not presently possible, due to computational/technological limitations.

\begin{table}

\centering

\caption{Specific penetration energy values.} 

\begin{tabular}{ccc} 

\hline 

Velocity (m/s) & Number of layers & Specific penetration energy (MJ/kg)\\ 

\hline
\hline

900 & 1  & 20.2 (MD)\\
1000 & 1  & 24.5 (MD)\\
1100 & 1  & 26.9 (MD)\\
2000 & 1  & 23.6 \cite{xia2016failure} (MD) \\
5000 & 1  & 29.0 \cite{yoon2016atomistic} (MD)\\
5000 & 1  & 40.8 \cite{haque2016molecular} (MD)\\
\hline
\hline
900 & 2 & 9.9 (MD)\\
1100 & 2 & 15.0 (MD)\\
5000 & 2  & 25.2 \cite{haque2016molecular} (MD)\\
\hline
\hline
600 & 127 (average) & 1.09\cite{Lee2014} (EXP)\\
900 & 154 (average) & 1.26\cite{Lee2014} (EXP)\\
\hline
\end{tabular}
\label{tab}
\end{table}

\subsection*{Scaling law}

The decrease in specific penetration energy for the double layered case suggests this quantity might be a function of the number of layers, and that some size-effect rescaling is needed in order to directly contrast numerical and experimental results. Note this effect can also be observed in the results provided by Haque \textit{et al.} \cite{haque2016molecular}. In order to investigate this possibility, we applied the scaling law proposed by Pugno \cite{Pugno2007} to correlate results across different size scales.

The key to understanding these results is that the strength of a material subject to nanoindentation or tensile tests has been, under fairly general assumptions, shown to be a function of the structural size \cite{Pugno2007}. For a material with $N$ layers it is possible to write its strength $\sigma_{N}$ as
\begin{equation}
\sigma_{N}=\sigma_{\infty} \sqrt{1+\frac{N_c}{N+\smash{N_{c}^{'}}}},
\label{sigman}
\end{equation}
where $\sigma_{\infty}$ is the strength of the bulk material, while $N_{c}$ and $N_{c}^{'}$ are critical values to be determined. These three quantities can be obtained from the numerical and experimental ballistic results.

One possible way to define the specific penetration energy of an $N$-layered material is
\begin{equation}
d_N=\frac{E}{\rho A_p N t},
\end{equation}
in which $A_p,\rho,N,t$ are respectively the projectile cross section area, density, number of layers and thickness of the single layer. This quantity can be related to the specific strength of the $N-$layered material ($\sigma_N$) \cite{pugno2006ice,carpinteri2002one}. See the Supplementary Information for more details on this procedure. Thus, we can write
\begin{equation}
d_N=\frac{\sigma_N}{\eta\rho},
\label{dn}
\end{equation}
in which $\eta$ is the ratio between the area of the projectile cross-section and the area of the damaged zone. This number is lower than one if an area larger than the cross-sectional area of the projectile is uniformly impacted. 

For instance, by using graphene density ($\rho\approx 2200 $kg . m$^{-3}$) and our $d_1$ and $d_2$ simulation values for $v=1100$m/s, we can derive $\sigma_1$ and $\sigma_2$ from Eq. \ref{dn}
\begin{eqnarray}
\sigma_1&=&d_1\eta\rho=59.2 \text{ GPa} \\
\sigma_2&=&d_2\eta\rho=33.0 \text{ GPa},
\end{eqnarray}
where we considered $\eta=1$.

\begin{figure}
\centerline{\includegraphics[width=0.9\linewidth]{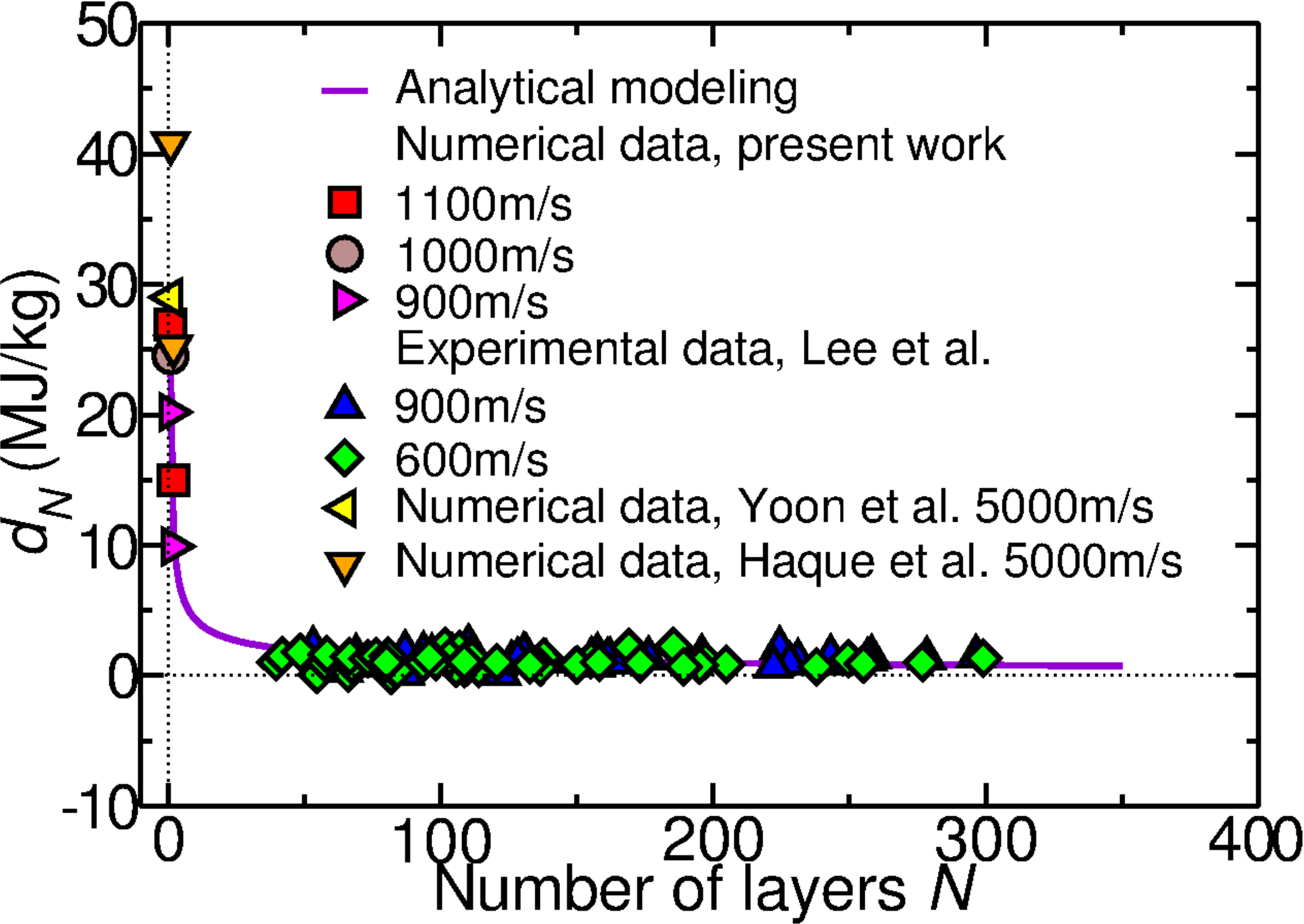}}
\caption{Analytical modeling fitting numerical results from molecular dynamics simulations carried out at the nanoscale and experimental ballistic test results carried out at the microscale by Lee \textit{et al.} \cite{Lee2014}.}
\label{energiaxmassa}
\end{figure}

A general expression for the specific penetration energy can be found from equations \ref{sigman} and \ref{dn},
\begin{equation}
d_N=d_\infty\sqrt{1+\frac{N_c}{N+\smash{N_{c}^{'}}}},
\label{dnfinal}
\end{equation}
where $d_\infty=\sigma_\infty/ \eta\rho$. 

We can fit previous \cite{Lee2014} and current results with equation \ref{dnfinal} to estimate the parameters $d_\infty$, $N_c$ and $N_{c}^{'}$.  Running a $100$ iterations best fit with tolerance $10^{-5}$ we found $d_\infty=0.26$, $N_c=2584$ and $N_{c}^{'}=-0.70$. After all parameters are obtained, we can use equation Eq. \ref{dnfinal} to estimate the specific penetration energy for any number of layers. The values obtained for few-layer graphene sheets are much higher than those obtained in the microscale (up to 30 times higher), suggesting a very sharp transition in the scaling law - see Fig. \ref{energiaxmassa}. Other theoretical results \cite{yoon2016atomistic,haque2016molecular} are also presented for comparison. Since the highest energy absorption per affected graphene mass is obtained when $N$ is small, thin graphene nanocoatings could be employed to maximize this quantity in ballistic applications.

\begin{figure}
\centerline{\includegraphics[width=0.8\linewidth]{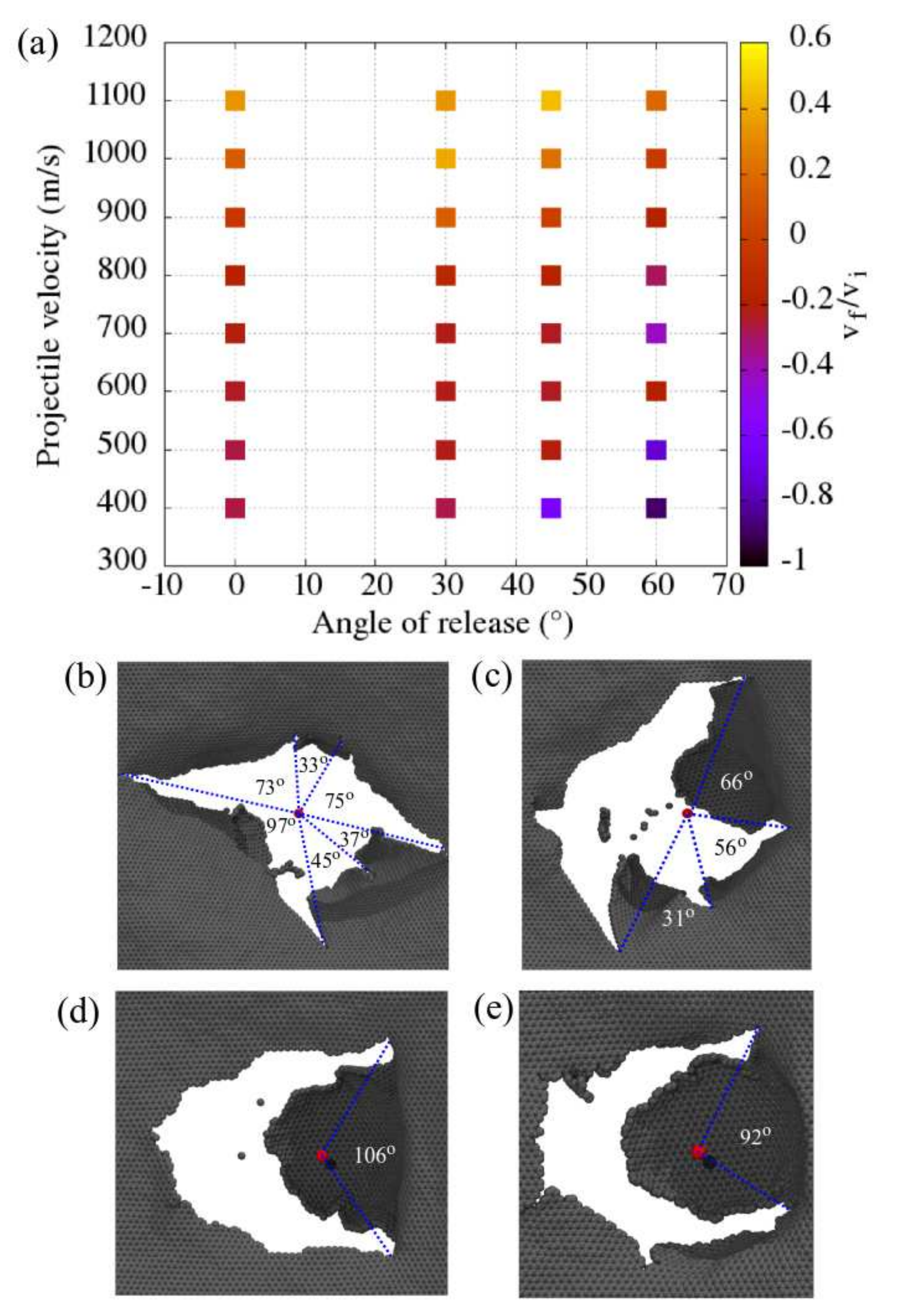}}
\caption{(a) Summary of results for the projectile kinetic energy value variations as a function of impact angle and initial velocity. Penetration occurs if the ratio $v_f$/$v_i$ is positive, otherwise the projectile returns. (b-d) Fracture patterns after impact for angle values of (b) $\theta=0^\circ$, (c) $\theta=30^\circ$, (d) $\theta=45^\circ$, and (e) $\theta=60^\circ$. Also indicated in these snapshots are the angle values between adjacent cracks.
}\label{angle}
\end{figure}

We also considered non perpendicular projectile impacts against single layer graphene sheets. Summary of the results for collisions with $\theta \neq 0^\circ$  are presented in Fig. \ref{angle}a, in which $v_i$ and $v_f$ are, respectively, the velocities before and after collision. Under these conditions we observed that collisions are rather elastic for higher impact angles and low velocities, in which the projectile can even bounce back. Penetration occurs whenever $v_f/v_i>0$. Inspection of Fig. \ref{angle}a reveals that for higher impact velocities penetration occurs regardless of the impact angle. Fracture patterns for different impact angles are presented in Fig. \ref{angle}b-e. As previously mentioned, our fracture patterns for $\theta=0^\circ$ are in good agreement with those reported by Lee \textit{et al.} \cite{Lee2014}, regarding both petal quantity and average opening angle between them. This suggests fracture patterns are scale independent, increasing the reliability of the predicted fracture patterns presented for alternate impact angles in Fig. \ref{angle}c-e. 

\section*{Conclusions}

In summary, we combined MD simulations and analytical modeling to explain the apparent discrepancies between numerical and experimental results for the specific penetration energy of graphene under ballistic impact. In the MD part of this work, we shot nickel projectiles at varied angles and velocities against single and double-layer graphene sheets, and studied the resulting dynamics and fracture patterns. Our results for perpendicular impacts were in good agreement with experimental data, suggesting these patterns are scale independent. The values we obtained for specific penetration energy from these simulations were consistent with previous numerical reports for single-layer graphene \cite{yoon2016atomistic}, but were an order of magnitude greater than experimental values for multi-layer sheets \cite{Lee2014}. Our analytical model suggests this disparity is due to size-scale effects, and the proposed power law was able to produce an excellent fitting of the numerical and experimental results obtained in different scale regimes. If extrapolated to the macroscopic limit, the model predicts a specific penetration energy of $~0.26$ MJ/kg. Although this macroscale value is still comparable to that of other ballistic protection materials \cite{Lee2014}, our results suggest that superior gravimetric performance in ballistic applications can be obtained by applying thin graphene nanocoatings over other materials.

\section*{Methods}

\subsection*{Computational Methods and details}

Our molecular dynamics (MD) simulations were carried out using the Reactive Force Field (ReaxFF) \cite{reax,Aktulga12}, as implemented in the LAMMPS software package \cite{lammps}. We used the parametrization described in Mueller \textit{et al.} \cite{Mueller2010}. ReaxFF is a reactive force field parametrized using ab-initio methods. It allows for the formation and dissociation of chemical bonds, making it potentially applicable to simulation of fractures at the nanoscale.

As the projectile, we used a $14000$ atom nickel nanoparticle, packed into a $\sim 3.5$ nm radius sphere. As the target, we used periodic graphene sheets, ranging from $20$ nm $\times$ $20$ nm ($30000$ atoms) to $100$ nm $\times$ $100$ nm ($385000$ atoms). We employed the following procedure in our simulations:
\begin{enumerate}
\item We minimized and thermalized the nickel nanoparticle for $200$ ps at $300$ K in the $NVT$ ensemble 
\item We minimized and thermalized the graphene sheet for $200$ ps at $300$ K in the $NPT$ ensemble. To reduce the initial stress, we set a null pressure at the edges of the structure
\item We thermalized the graphene sheet for an additional $200$ ps at $300$ K in the $NVT$ ensemble
\item We fixed the edges of the graphene unit cell, to prevent uniform translation of the sheet during impact
\item We shot the projectile against the graphene sheet in the $NVE$ ensemble, with velocity \textbf{v} and angle $\theta$. Different \textbf{v} and $\theta$ values were considered.
\end{enumerate}
For steps 1 to 3 we used a timestep of $0.5$ fs while in step 5 we used a timestep of $0.02$ fs. Temperature and pressure were controlled through chains of three Nos\'e-Hoover thermostats and barostats \cite{shinoda2004rapid}.

\subsection*{Procedure to calculate the specific penetration energy from the data published by Haque \textit{et al.} \cite{haque2016molecular} }
In that paper, the energy transferred to the graphene sheet during a ballistic test is $E_T^{GS}$. In order to obtain the specific penetration energy ($d_N$), this energy has to be divided by the graphene mass within the projectile cross section. This mass is equal to  $m= \pi R^2 N_{L} \rho_A$, where $R$ is the projectile radius, $N_L$ is the number of layers, and $\rho_A=0.77$ mg/m$^2$ is the area density of graphene. For $v=5000$ m/s, we obtained from the manuscript that  $E_T^{GS}=36.13$ aJ for $N_L=1$ and $E_T^{GS}=44.65$ aJ for $N_L=2$.  After dividing these results by the mass, we get $d_1= 40.8$ MJ and $d_2=25.2$ MJ. More $E_T^{GS}$ data is presented in the paper, but this is the only velocity for which results are presented in which complete penetration is observed for different number of layers.

\subsection*{Procedure to extract data from Lee \textit{et al.} \cite{Lee2014}, Yoon \textit{et al.} \cite{yoon2016atomistic}, and Xia \textit{et al.} \cite{xia2016failure}}
In order to extract data from figure 4c of Lee \textit{et al.} \cite{Lee2014}, figure 4a of Yoon \textit{et al.} \cite{yoon2016atomistic}, and figure 9a of Xia \textit{et al.} \cite{xia2016failure}, we used the web app WebPlotDigitizer \cite{website}.


\section*{Acknowledgements}

This work was supported by CAPES, CNPq, FAPESP, ERC and the Graphene FET Flagship.
RAB, LDM, JMS and DSG would like to thank the Center for Computational Engineering and Sciences at Unicamp for financial support through the FAPESP/CEPID Grant $2013/08293-7$. 
LDM acknowledges financial support from the Brazilian Federal Agency CAPES \textit{via} its PNPD program. NMP is supported by the European Research Council PoC 2015 ``Silkene" No. 693670, by the European Commission H2020 under the Graphene Flagship Core 1 No. 696656 (WP14 ``Polymer Nanocomposites") and under the Fet Proactive ``Neurofibres" No. 732344.

\section*{Author contributions statement}

D.S.G, R.A.B., and L.D.M. conceived the protocols for the molecular dynamics simulations, and R.A.B. and L.D.M. performed the simulations. R.A.B, L.D.M, and J.M.S analysed the simulation results. N.M.P. proposed the analytical modeling and supervised its execution by R.A.B. R.A.B and L.D.M. wrote the manuscript. All authors discussed and revised the manuscript. 

\section*{Additional information}

\textbf{Competing financial interests}: The authors declare no competing financial interests. 


\end{document}